\documentclass{svmult}

\usepackage{makeidx}         % allows index generation
\usepackage{graphicx}        % standard LaTeX graphics tool
\usepackage{multicol}        % used for the two-column index
\usepackage[bottom]{footmisc}% places footnotes at page bottom

\makeindex             % used for the subject index
\begin{document}

%----------------------------------------------------------------------
%  TITLE AND AUTHOR LIST
%----------------------------------------------------------------------
\title*{Investment horizons : A time-dependent measure of asset
  performance}

\author{Ingve Simonsen\inst{1}\and Anders Johansen\inst{2} 
             \and  Mogens H.\  Jensen\inst{3}}

\institute{Department of Physics, NTNU, NO-7491 Trondheim, Norway 
%\texttt{ingves@phys.ntnu.no}
\and 
Teglg\aa rdsvej 119, DK-3050 Humleb\ae k, Denmark 
%\texttt{anders-johansen@get2net.dk}
\and
Niels Bohr Institute, Blegdamsvej 17, DK-2100 Copenhagen {\O}, Denmark 
%\texttt{mhjensen@nbi.dk}
}

\date{\today}
\maketitle

%----------------------------------------------------------------------
%  ABSTRACT AND KEYWORDS
%----------------------------------------------------------------------

\begin{abstract}
  We review a resent {\em time-dependent} performance measure for
  economical time series --- the (optimal) investment horizon
  approach. 
  For stock indices, the approach
  shows a pronounced gain-loss asymmetry that is {\em not} observed for the
  individual stocks that comprise the index. This difference may hint
  towards an synchronize of the draw downs of the stocks.
\end{abstract}

% --------------------------------------------------------------------
%   MAIN TEXT
% --------------------------------------------------------------------

%-------------------------------------
%\section{Introduction}
%-------------------------------------

As an investor or practitioner working in the financial industry, you
are continuously faced with the challenge of how to chose and manage a
portefolio under varying market conditions; as the market change, you
have to decide whether to adjust your positions in order to make the
portfolio, as you see it, more optimal. The way such important
decisions are made, with dramatic economic consequences if done badly,
is rather complex; most market players have their very own methods for
this purpose, and they are only rarely disclosed to the public.  The
clients risk aversion, which is based on individual psychology, plays
a major role in the task of choosing a portfolio and hence
quantifiable and rational measure must be used in for example stress
testing of the portfolio. 

As the financial industry became fully computerized, the distribution
of returns approach became popular for measuring asset performance
from historic data records. Today, this method is considered one of
the classic approaches for gauging the performance of an
asset~\cite{Book:Bouchaud-2000,Book:Mantegna-2000}. The method relies
on the distribution of returns~(to be defined below) corresponding to
a {\em fixed} time window (or horizon as we will refer to it below).
In order to look into the performance over a different time horizon,
the return distribution has to be regenerated for the new window size.
Actually, one of the most successful strategies for actively investing
when the risk aversion is not low, is to, {\it a priori}, decide for a
return level and then liquidate the position when this level has been
reached.

It is not at all clear that the natural scenario for an investor is to
consider fixed time windows. There has therefore lately been a lot of
interest in time dependent measures, {\it i.e.}  measures where the
time period over which the asset is hold, is non-constant, and allowed
to depend on the specific market conditions which in general is not
known in detail. A change in time horizon used by an investor may be
due to for instance a changes in the market itself, or new investment
strategies being implemented by the investor. 

In this work, we will review a recent development in such
time-dependent measures --- the {\em investment horizon approach}. 
This approach is motivated by progress in turbulence~\cite{Mogens},
and it represents an adaption of a more general concept, known as {\em
  inverse statistics}, to economics. The investment horizon approach
was first introduced into economics by the present
authors~\cite{Simonsen2001-5}, and later considered in a series of
publications~\cite{Simonsen2002-7,Simonsen2002-6,Simonsen2004-1,WorkInProgress}. 
The method has recently been applied to different types of financial
data with success; stock index
data~\cite{Simonsen2001-5,Simonsen2002-7,Simonsen2002-6}, like the Dow
Jones Industrial Average~(DJIA), NASDAQ, Standard and Poor
500~(SP500), individual stocks~\cite{WorkInProgress}, and high
frequency foreign exchange~(FX) data~\cite{Simonsen2002-6}.  A similar
approach, however without a fixed return level, has been studied in
Refs.~\cite{Johansen-1998,Johansen-2001} with the prime focus on
losses.

%-------------------------------------
%\section{The Method}
%-------------------------------------
%\bigskip

Let $S(t)$ denote the asset price, and $s(t) = \ln S(t)$ the
corresponding logarithmic price. Here time ($t$) can be measured
in different ways~\cite{Book:Mantegna-2000}, but the various choices
may result in different properties for the inverse
statistics~\cite{Simonsen2002-6}.  The logarithmic return at time $t$,
calculated over a time interval $\Delta t$, is defined
as~\cite{Book:Bouchaud-2000,Book:Mantegna-2000} $r_{\Delta t}(t) =
s(t+\Delta t)- s(t)$. 

\begin{figure}[t]
    \begin{center}
        \leavevmode
        \includegraphics*[width=0.8\columnwidth,height=0.55\columnwidth]{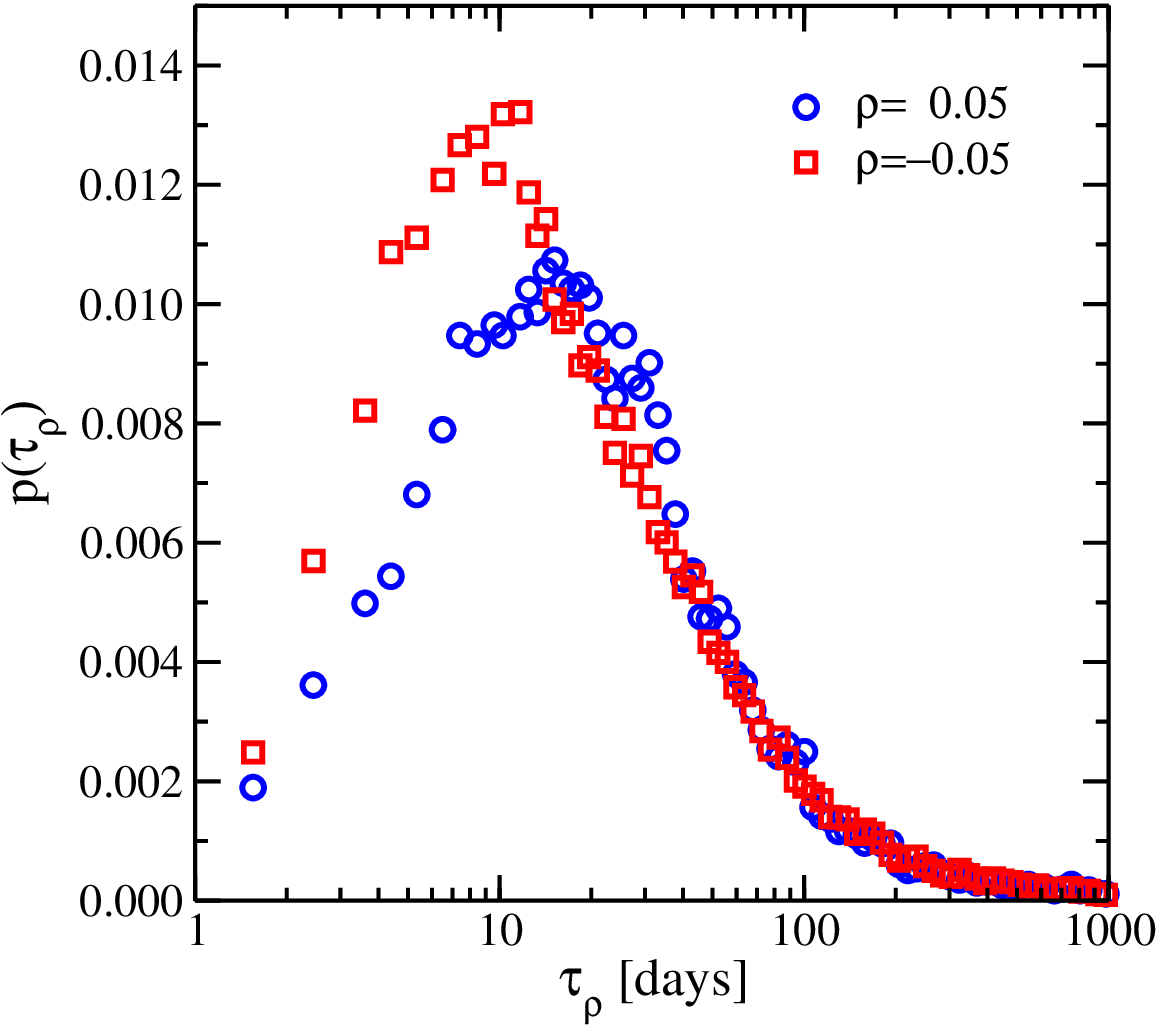}\\*[0.3cm]
        \includegraphics*[width=0.8\columnwidth,height=0.55\columnwidth]{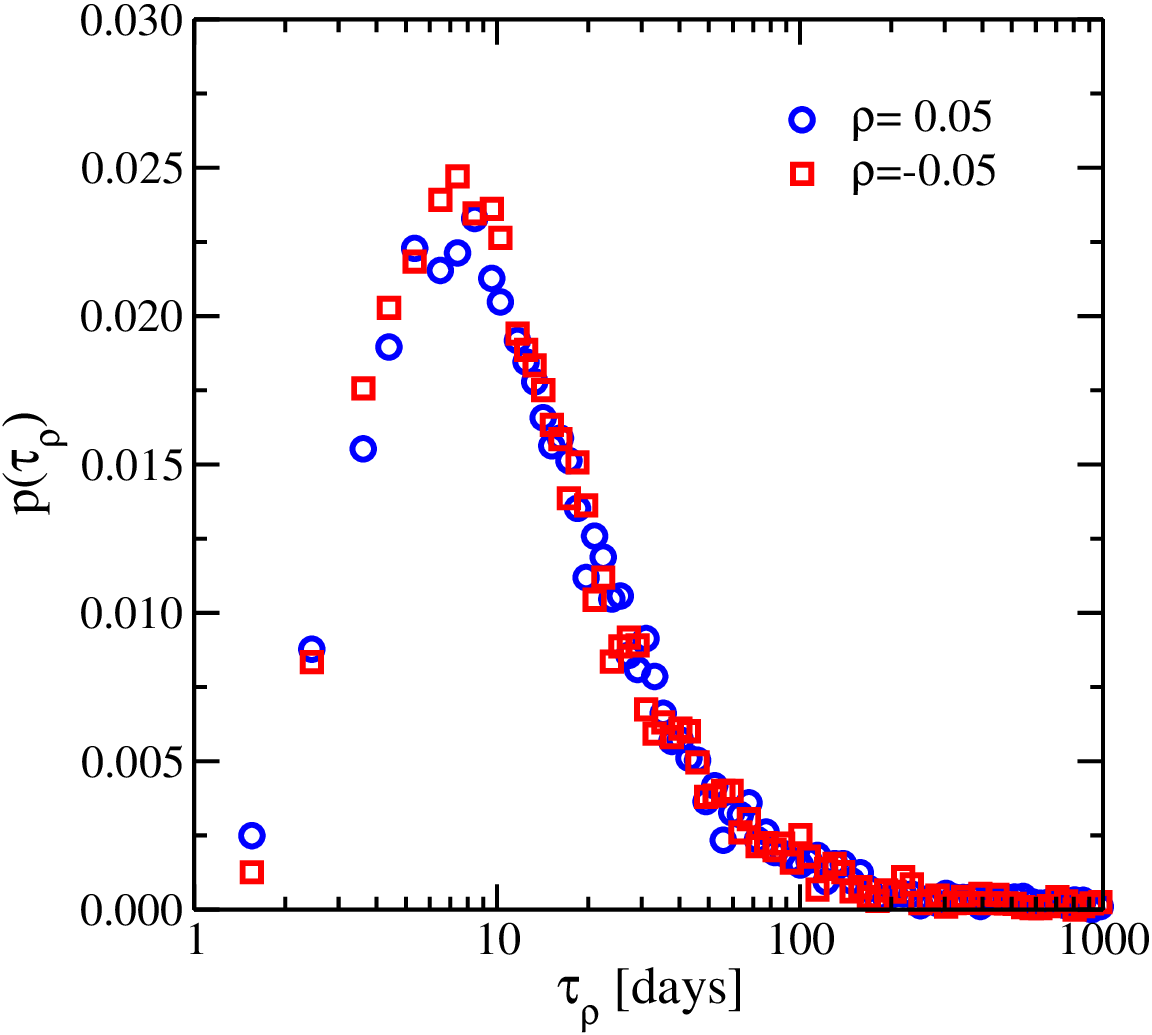}
      \begin{picture}(0,0)(0,0)
        \put(-200,350){\makebox(0,0)[b]{\large$(a)$}}
        \put(-200,150){\makebox(0,0)[b]{\large$(b)$}}
     \end{picture}
       \caption{(a) The investment horizon distributions of the DJIA
         closing prices from 1896 till present,  at a return levels
         $\left|\rho\right|=0.05$. 
         (b) The same as Fig.\protect\ref{Fig:DJIA_IBM_PDF}(a), but now
         for the single stock of IBM for the period from  the beginning of
         1962 till June 2000. IBM has been part of DJIA since June 29,
         1979. 
       }
        \label{Fig:DJIA_IBM_PDF}
    \end{center}
\end{figure}
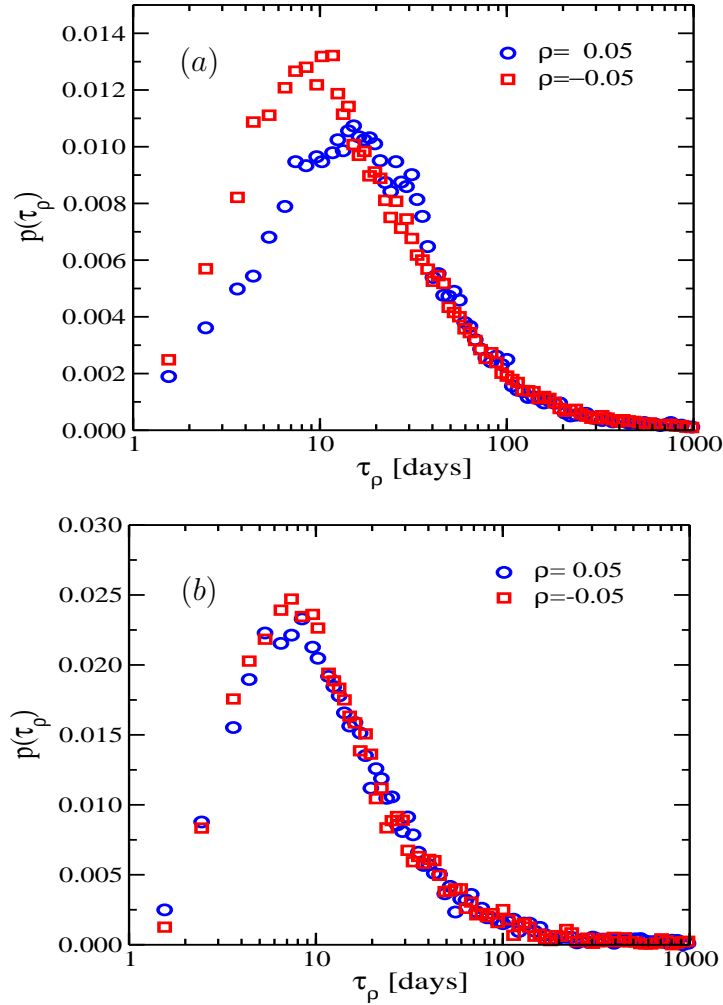

We consider a situation where an investor is aiming for a given return
level denoted by $\rho$. This level may be both positive (gains) or
negative (losses).  If the investment is made at time $t$, then the
investment horizon is defined as the time $\tau_\rho(t)=\Delta t$ so
that the inequality $r_{\Delta t}(t)\geq \rho$ ($r_{\Delta t}(t)\leq
\rho$) for $\rho\geq 0$ ($\rho<0$) is satisfied for the {\em first}
time.  In mathematical terms, this can be expressed as
\begin{eqnarray}
  \label{InvestmentHorizon}
  \tau_\rho(t) &=& \left\{  
    \begin{array}[]{ll}
        \inf \left\{ \Delta t \; | \; r_{\Delta t}(t) \geq \rho
          \right\},
               & \quad \rho \geq 0, \\    
        \inf \left\{ \Delta t \; | \; r_{\Delta t}(t) \leq \rho 
           \right\},
               & \quad \rho < 0.    
     \end{array}
   \right. 
\end{eqnarray}
The investment horizon distribution, $p\left( \tau_\rho\right)$, is
then the distribution of investment horizons $\tau_\rho$ estimated
from the data (cf. Fig.~\ref{Fig:DJIA_IBM_PDF}a). This distribution
will go through a maximum, as should be apparent from the discussion
to follow.  This maximum --- the {\em optimal investment horizon} ---
will be denoted $\tau_{\rho}^*$. It quantifies the most likely time
period (obtained from historic data) needed to reach the investment
outcome characterized by $\rho$.

For later use, we stress that if the price process $S(t)$ is a
geometrical Brownian motion --- the classic assumption made in
theoretical finance --- then the solution to the investment horizon
(first passage time) problem is known
analytically~\cite{Book:Kannan-1979}. It can be shown that the
investment horizon distribution is given by the Gamma-distribution:
$p(t) = \left|a\right|\exp(-a^2/2t)/(\sqrt{2\pi}t^{3/2})$, where
$a\propto \rho$. Hence, in the limit of large (waiting) times, one
recovers the well-known first return probability $p(t) \sim t^{-3/2}$.

%-------------------------------------
%\section{The Analysis}
%-------------------------------------
%\bigskip

%---------------------------------------------------------
%---------------------------------------------------------

\begin{figure}[t]
    \begin{center}
      \leavevmode 
       \includegraphics*[width=0.8\columnwidth,height=0.55\columnwidth]{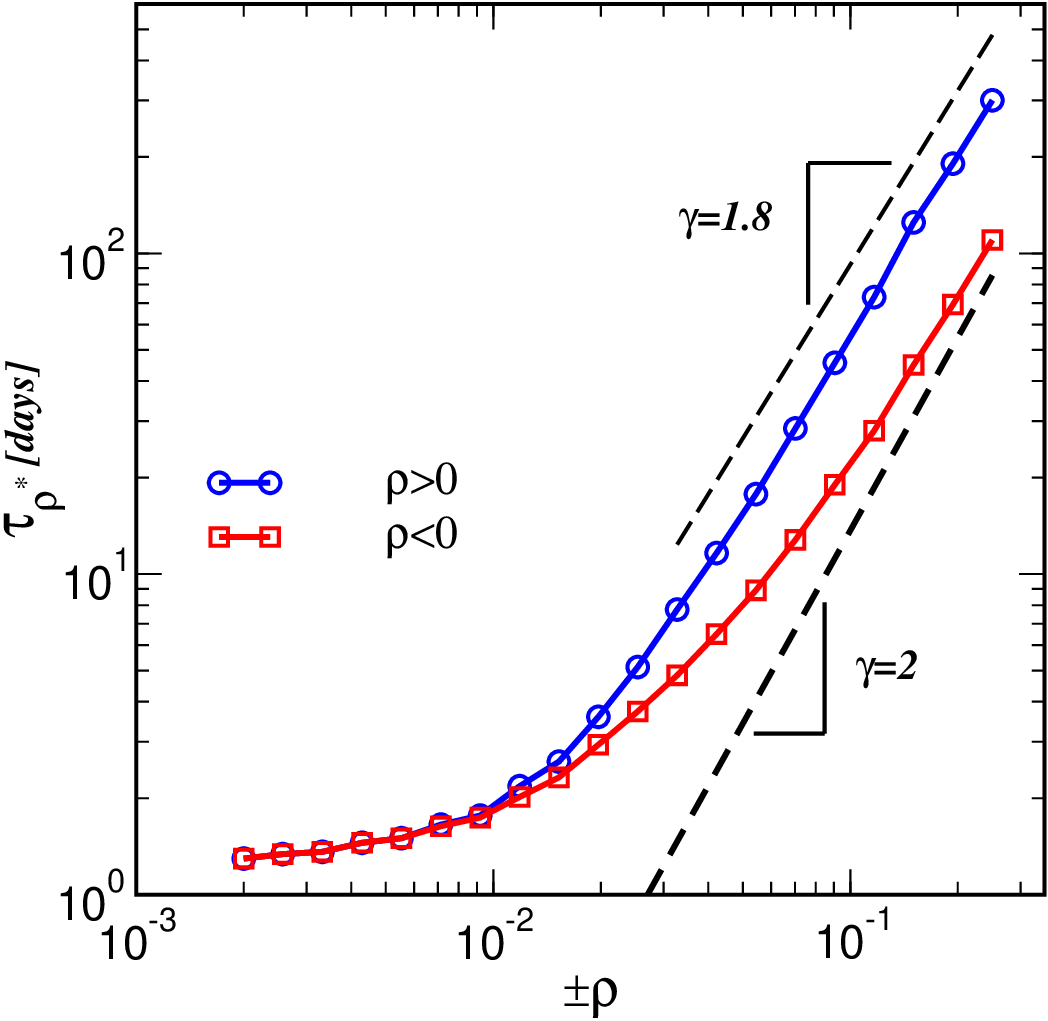}
       \caption{The optimal investment horizon $\tau^*_\rho$ for
         positive~(open circles) and negative~(open squares) levels of
         return $\pm \rho$. In the case $\rho<0$ one has used
         $-\rho$ on the abscissa for reasons of comparison.} 
        \label{Fig:DJIA_PDF_scaling}
    \end{center}
\end{figure}

Figures~\ref{Fig:DJIA_IBM_PDF} show empirical investment horizon
distributions, $p\left( \tau_\rho\right)$ with $\rho=\pm 0.05$. for an
index (Fig.~\ref{Fig:DJIA_IBM_PDF}a) and an individual stock
(Fig.~\ref{Fig:DJIA_IBM_PDF}b).  Drift-terms that were ``smooth'' up
till a time scale of roughly $4$ years, were removed from the
logarithmic prices prior to the analysis (consult
Ref.~\cite{Simonsen2001-5} for details).  This pre-processing of the
data was done in order to enable a more consistent comparison of the
results corresponding to positive and negative levels of
returns due to differences in economic fundamentals
such as inflation, interest rates, {\it etc.} 
The data set used to produce the results of
Fig.~\ref{Fig:DJIA_IBM_PDF}a was the daily close of the Dow Jones
Industrial Average taken over its whole history up till present.  From
this same figure, two well-pronounced, but not coinciding, optimal
investment horizons can be observed from the empirical distributions
$p\left(\tau_\rho\right)$.  With $|\rho|=0.05$ they are both of the
order of $\tau_\rho^*\sim 10$~days.  In general, the values of
$\tau_\rho^*$ will depend on the return level $\rho$, and we presents
results for the DJIA in Fig.~\ref{Fig:DJIA_PDF_scaling} for positive
and negative return levels.
Recall that if the price process is consistent with a geometrical
Brownian motion, one has $\tau^*_\rho\sim \rho^\gamma$ with $\gamma=2$
for {\em all} values of $\rho$ (lower dashed line in
Fig.~\ref{Fig:DJIA_PDF_scaling}). The empirical results are observed
not to be consistent with such a behavior. For rather small levels of
returns --- a fraction of a percent --- the dependence on return level
is quite weak. When $|\rho|$ is increased, however, the dependence
becomes more pronounced and it gradually becomes more and more
like, but still different from, the geometrical Brownian result. As a
whole, the dependence of $\tau^*_\rho$ (on $\rho$) over the range of
return levels considered in Fig.~\ref{Fig:DJIA_PDF_scaling}, resembles
more a double logarithmic behavior than a power law.  However, for the
range of $\rho$-values considered and the fact that the statistics
become poorer for increasing levels of return, we are unable from the
empirical data alone to uncover the actual functional dependence of
$\tau^*_\rho$ on the return level.

%---------------------------------------------------------

%---------------------------------------------------------

One of the most striking features of Fig.~\ref{Fig:DJIA_PDF_scaling}
is the apparent fact that the optimal investment horizon for positive
and negative return levels are {\em not} the same. This asymmetry
starts to develop when the return level $|\rho|$ is not too small (cf. 
Fig.~\ref{Fig:DJIA_PDF_scaling}). Such a {\em gain-loss asymmetry} is
actually a rather general feature of the investment horizon of stock
indices~\cite{Simonsen2002-6}.  
On the other hand, for individual stocks that together comprise
the index, this phenomenon is less pronounced~\cite{WorkInProgress}
and an asymmetry can often hardly be seen at all. In
Fig.~\ref{Fig:DJIA_IBM_PDF}b this is exemplified by the investment
horizons of IBM for $\rho=\pm 0.05$, a company that is part of the DJIA
index. 
Similar results hold for most other stocks being part of the
DJIA~\cite{WorkInProgress}.  The attentive reader could ask: How is it
possible that an asymmetry is present in the index, but not in the
individual stocks that {\em together} make out the index? At the time
of writing, there is no consensus on what is causing this behavior. It
has been speculated that it might be caused by cooperative effects
taking place among the stocks and causing them to partly synchronize
their draw-downs (conditional serial correlation). If that was to be
the case, the index --- that is some average of the individual stocks
--- will experience an increased probability of shorter investments
horizons $\tau_{-|\rho|}$ compared to the similar results found for
the same positive level of return.  Other speculations go in the
direction of this phenomenon being related to the so-called leverage
effect~\cite{leverage}.  These questions are being addressed by
ongoing research efforts, and it is hoped that they will be
satisfactory answered in the immediate future. 

%\bigskip 
Before ending this contribution, we would like to add a few comments
regarding possible practical implications (as we see it) of the
investment horizon approach~\cite{Ted}. Two applications will be
mentioned here, both taken from portfolio management. The first
application is related to the problem of consistent allocation of
VAR-like (quantile) and stop-loss limits. For such problems, the
correlation structure over different time horizons is important. Our
approach naturally use non-fixed time windows, and it is therefore
hoped that it might contribute some new insight onto these issues.
The second application is concerned with the calculation of risk
measures for portfolios.  When the market is moving against you, you
are forced to liquidate.  In this process, ``liquidation horizons''
that are used across assets of a portfolio, are normally not the same.
By taking advantage of the negative return levels, investment horizon
distributions $p\left(\tau_{-|\rho|}\right)$ for the different assets
of the portfolio, may be used to design an {\em optimal liquidation}
procedure depending on the nature of the position, {\it e.g.}, long or
short.  The exploration of possible applications of the concept of
inverse statistics in economics is at its infancy. We hope that the
future will demonstrate this approach to be fruitful also from a
practical standpoint.

%-------------------------------------
%\section{Conclusions and outlook}
%-------------------------------------
\smallskip

A new measure of asset performance that represents an alternative to
the classic distribution of returns approach has been described. 
Unlike the classic method, the new technique is {\em time-dependent}. 
This opens the possibility of studying and measure asset performance
over a non-constant time scale, an idea that lately has attracted a
great deal of attention. 

\bigskip

%-------------------------------------
%\section*{Acknowledgements}
%-------------------------------------
\noindent
{\bf Acknowledgements}\\
The first author wishes to thank Drs. Ted Theodosopoulos and Marc
Potters for fruitful discussions and valuable comments and
suggestions.  IS also acknowledges the financial support kindly
provided by Nihon Keizai Shimbun Inc.

% --------------------------------------------------------------------
% BIBLIOGRAPHY
% --------------------------------------------------------------------
\bibliographystyle{mybibstyle}
\bibliography{publication_list,books,paper2005-1}

\printindex
\end{document}